\documentclass[prl,twocolumn,floatix,amssymb,amsmath,superscriptaddress,letterpaper]{revtex4}
\usepackage{hyperref}
\usepackage{graphicx}
\usepackage{color}
\usepackage{multibbl}

\begin{document}
\title{Experimental Demonstration of Robust Bidirectional Quantum Optical Communications}

\author{Jin-Shi Xu}
\thanks{These authors contributed equally to this work.}
\affiliation{Key Laboratory of Quantum Information, University of Science and Technology of China, CAS, Hefei, 230026, People's Republic of China}

\author{Man-Hong Yung}
\thanks{These authors contributed equally to this work.}
\affiliation{Department of Chemistry and Chemical Biology, Harvard University, Cambridge MA, 02138, USA}
\affiliation{Center for Quantum Information, Institute for Interdisciplinary Information Sciences, Tsinghua University, Beijing, 10084, People's Republic of China}

\author{Xiao-Ye Xu}
\affiliation{Key Laboratory of Quantum Information, University of Science and Technology of China, CAS, Hefei, 230026, People's Republic of China}

\author{Jian-Shun Tang}
\affiliation{Key Laboratory of Quantum Information, University of Science and Technology of China, CAS, Hefei, 230026, People's Republic of China}

\author{Chuan-Feng~Li}
\email{cfli@ustc.edu.cn}
\affiliation{Key Laboratory of Quantum Information, University of Science and Technology of China, CAS, Hefei, 230026, People's Republic of China}

\author{Guang-Can Guo}
\affiliation{Key Laboratory of Quantum Information, University of Science and Technology of China, CAS, Hefei, 230026, People's Republic of China}

\date{\today }

\begin{abstract}
We experimentally realized a new method for transmitting quantum information reliably through paired optical polarization-maintaining (PM) fibers.
The physical setup extends the use of a Mach-Zehnder interferometer, where noises are canceled through interference.
This method can be viewed as an improved version of the current decohernce-free subspace (DFS) approach in fiber optics.
Furthermore, the setup can be applied bidirectionally, which means that robust quantum communication can be achieved from both ends.
To rigorously quantify the amount of quantum information transferred, optical fibers are analyzed with the tools developed in quantum communication theory. These results not only suggests a practical means for protecting classical and quantum information through optical fibers, but also provides a new physical platform for enriching the structure of the quantum communication theory.
\end{abstract}


\maketitle

One of the most well-known methods for achieving robust quantum communication is the idea of decoherence-free subspace (DFS)~\cite{Lidar2003}, when quantum states are attacked by correlated errors. It has been shown that a DFS exists when transmitting quantum information down a dephasing channel with memory~\cite{DArrigo2007} and entangled states are used to enhance classical communication over a channel with correlated noise~\cite{Banaszek2004}.
For optical-fiber communication, DFS have been experimentally applied to generation of entanglement pairs \cite{Brendel1999,Marcikic2004,Honjo2008,Yamamoto2008} and direct quantum-state transfer \cite{Chen2006,Yamamoto2007}. In spite of its elegance, the current DFS approach for optical communications suffers from drawbacks that make it challenging to achieve large-scale applications. First, the resources for obtaining deterministic entangled photons or achieving entangling operations are still demanding (with low efficiency) for the current quantum-optics technology \cite{Pan2012}. Second, for direct transfer of quantum states, many optical implementations of DFS \cite{Chen2006,Yamamoto2007} produce correct output states only probabilistically (i.e., post-selection), which limits the potential applicability of the methods for multiple uses.
Third, existing approaches \cite{Banaszek2004,Brendel1999,Marcikic2004,Honjo2008,Yamamoto2008,Chen2006,Yamamoto2007} for quantum communication through fiber optics are uni-directional (i.e. one-way communication)~\cite{comment}, which limits the range of theirs applications.

Here we propose a new approach of protecting quantum information through noisy optical fibers, which can be operated in both unidirectional and bidirectional modes. Bidirectional quantum communication is important for many applications including two-way quantum key distribution~\cite{Cere2006} and realizations of quantum interactive proof systems~\cite{Watrous2003}. The key feature of our method compared with the current DFS approach is that when DFS is applied to communication channels of optical fibers~\cite{Banaszek2004,Chen2006,Yamamoto2008}, quantum information is typically encoded with a pair of entangled photons,
which are successively sent through the same optical fiber (i.e., 2 photons + 1 fiber). Here we consider a more versatile implementation that protects individual photons based on the inference of their paths, i.e., a single-photon phenomenon (1 photon + 2 fibers, and $n$ photons + 2$n$ fibers). This difference is crucial, as efficient means for generating entangled photons on demand are currently challenging to achieve in the laboratory.

Distinct from the other related experiments \cite{Banaszek2004,Brendel1999,Marcikic2004,Honjo2008,Yamamoto2008,Chen2006,Yamamoto2007}, we performed our study with the tools developed quantum communication theory. The key difference is that we focus on the amount of quantum information that can transmitted through the channels, apart from measuring the fidelity of the transmitted states. These results allow us to predict the quality of transmitted states for any input state. Furthermore, as far as we know, this is the first time in the literature where the quantum capacities of optical fibers are systematically analyzed in the laboratory, which establishes a connection between theoretical quantum communication theory and experimental quantum optics.

\begin{figure*}[t]
\begin{center}
\includegraphics [width= 6 in]{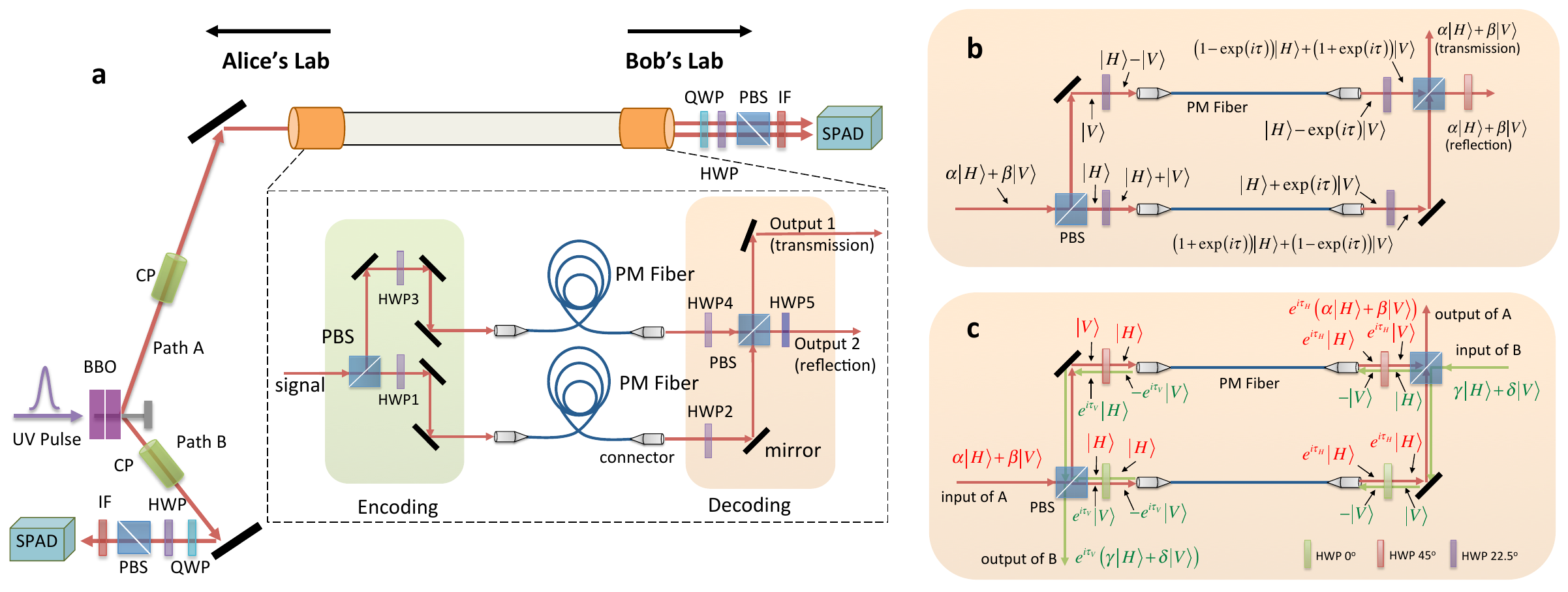}
\end{center}
\caption{(Color online). {\bf a}, The full setup for entanglement distribution over a pair of 120-meter-long polarization-maintaining (PM) fibers. Part of the entangled photon is kept by Alice's lab; another part of the entangle pair enters the interferometric unit. The photons are finally detected by single photon avalanche detectors (SPADs) with 3 nm interference filters (IFs) in front of them. {\bf b}, The quantum state evolution of a photon in the case where the noise from the two fibers are perfectly correlated. In the experiment, the correlation is only partial. {\bf c}, The state evolution of the bi-directional information transfer of the setup.
} \label{fig:setup}
\end{figure*}

{\em Experimental setup--- }Our experimental setup is shown in Fig.~\ref{fig:setup}a. The key ingredients in our setup include a pair of optical polarization-maintaining (PM) fibers and a set of half-wave plates (HWP). The two types of setting of the HWPs allow the setup to be operated in the unidirectional (Fig.~\ref{fig:setup}b) or bidirectional (Fig.~\ref{fig:setup}c) mode. The efficiencies of both modes were systematically analyzed in our experiment.

The main source of errors in PM fibers come from optical birefringence~\cite{Kumar2011}, where the relative-phase fluctuation (or dephasing) between the pair of polarization (horizontal $\left| H \right\rangle$ and vertical $\left| V \right\rangle$) depend on the frequency $\omega$ of the photon.
In the experiment, the PM fibers are physically bundled together in order to maximize noise correlation.

{\em Unidirectional mode---} A pictorial illustration of the unidirectional mode is shown in Fig. \ref{fig:setup}b for the idealized case~\cite{SM}. The theoretical description of the evolution for a signal photon, with frequency $\omega$, carrying one qubit of information, $\alpha|H\rangle+\beta|V\rangle$ ($\alpha$ and $\beta$ are some unknown complex numbers and $|\alpha|^2+|\beta|^2=1$), is as follows: the photon is sent through the first polarization beam splitter (PBS), and gets decomposed into two paths, $\alpha|H\rangle |0\rangle+\beta|V\rangle |1\rangle$. Here $|0\rangle$ ($|1\rangle$) denotes the transmitted (relfected) part. After passing through half-wave plates (HWP1 and HWP3) with angles set to be $22.5^{\circ}$, the state becomes $\alpha \left|  D  \right\rangle \left| 0 \right\rangle  {+} \beta \left|  J  \right\rangle \left| 1 \right\rangle$, where $|D\rangle=\frac{1}{\sqrt{2}}(|H\rangle+|V\rangle)$ and $|J\rangle=\frac{1}{\sqrt{2}}(|H\rangle-|V\rangle)$. Then, these two components are directed to the PM fibers each of length 120 m. For each frequency mode, the resulting state is of the form: $\frac{\alpha}{\sqrt{2}} ( {\left| H \right\rangle  + {e^{i{\tau _1}\left( \omega  \right)}}\left| V \right\rangle } ) \left| 0 \right\rangle  + \frac{\beta}{\sqrt{2}}( {\left| H \right\rangle  - {e^{i{\tau _2}\left( \omega  \right)}}\left| V \right\rangle } )\left| 1 \right\rangle$, where
\begin{equation}
\tau_j (\omega)=(L_j \Delta n_j /c) \omega \quad,
\end{equation}
$L_j$ is the fiber length,  $c$ is the vacuum light speed, and $\Delta n_j$ is the difference of the refractive indexes between the polarizations.

Note that there is a phase difference, $\Delta \tau  \equiv {\tau _1} - {\tau _2}$, between the two paths in general. After passing through the two other HWPs (HWP2 and HWP4) with the same angles set as in the encoding stage, we apply another PBS to allow the two paths to have optical inteference. At one of the outputs, we further apply a HWP with the angle setting at $45^{\circ}$, which flips between $|H\rangle$ and $|V\rangle$ ($|H\rangle\Leftrightarrow|V\rangle$).

The resulting state becomes~\cite{SM}
$ \left( {\alpha {\mu _{1 + }}\left| H \right\rangle  + \beta {\mu _{2 + }}\left| V \right\rangle } \right)\left| 0 \right\rangle  + \left( {\alpha {\mu _{1 - }}\left| H \right\rangle  + \beta {\mu _{2 - }}\left| V \right\rangle } \right)\left| 1 \right\rangle$, where ${\mu _{j \pm }} \equiv {\textstyle{1 \over 2}}\left( {1 \pm {e^{i{\tau _j}\left( \omega  \right)}}} \right)$. In the limit where the two fibers are nearly identical, i.e., ${\tau _1} \approx {\tau _2} \equiv \tau$, the protocol can perfectly transmit the input signal, $\alpha|H\rangle+\beta|V\rangle$, to either of the output ports, independent of any frequency $\omega$. In general, partial quantum information can still be transmitted through the fibers, as long as the noise are correlated.

{\em Bidirectional mode---} The physical setting for the bidirectional mode is shown in fig. \ref{fig:setup}c. When a quantum state, $\alpha|H\rangle+\beta |V\rangle$, is sent from the A side (from left to right), the PBS entangles the path DOF, which gives $\alpha |H\rangle \left| 0 \right\rangle  + \beta |V\rangle \left| 1 \right\rangle$. The polarization along the transmitted path $(\left| 0 \right\rangle )$ is unchanged, but the half-wave plate (HWP) located at the reflected path $(\left| 1 \right\rangle )$ flips the polarization from the vertical state to the horizontal state. The information of the input quantum state is then transferred to the path DOF, $|H\rangle \left( {\alpha \left| 0 \right\rangle  + \beta |\left| 1 \right\rangle } \right)$. After the photon passes through the pair of PM fibers, there are phase shifts ${{\tau_{H1,H2}}\left( \omega  \right)}$ induced from each path for each frequency component $\omega$, $\left| H \right\rangle \left( {\alpha {e^{i{\tau_{H1}}\left( \omega  \right)}}\left| 0 \right\rangle  + \beta {e^{i{\tau_{H2}}\left( \omega  \right)}}|\left| 1 \right\rangle } \right)$. Then, another HWP flips the polarization back to $|V\rangle $, which gives the following state $\alpha {e^{i{\tau_{H1}}\left( \omega  \right)}}\left| H \right\rangle \left| 0 \right\rangle  + \beta {e^{i{\tau_{H2}}\left( \omega  \right)}}\left| V \right\rangle |\left| 1 \right\rangle $. Finally, the two paths cross at another PBS, which separates the path DOF, $\left( {\alpha {e^{i{\tau_{H1}}\left( \omega  \right)}}\left| H \right\rangle  + \beta {e^{i{\tau_{H2}}\left( \omega  \right)}}\left| V \right\rangle } \right)\left| 0 \right\rangle$. Whenever the errors are correlated, i.e., ${\tau_{H1}} \approx {\tau_{H2}} \equiv \tau_{H}$, the signal photon can be detected along the path $\left| 0 \right\rangle$. The phase difference ${\tau_{H1}}\left( \omega  \right) - {\tau_{H2}}\left( \omega  \right)$ causes dephasing to the output state as expected. In a similar way, quantum information can be sent from the opposite direction (from right to left, i.e., from B to A) as indicated in Fig. \ref{fig:setup}c~\cite{SM}.

{\em Quantifying quantum-information transferred---} In order to quantify the amount of quantum information that can be transmitted through our setup, we analyze the data through the tools developed in quantum communication theory, instead of just measuring fidelity of output states. However, as far as we are aware of, the parameterization in this work is not widely known; therefore, we provide a concise but self-contained theoretical background below~\cite{SM}.

Quantitatively, the amount of quantum information transmitted through a noisy channel can be quantified by {\it coherent information}~\cite{Schumacher1996},
\begin{equation}
\mathcal{I}_{c}=S[\mathcal{E}(\rho_{A})]-S[(\mathcal{E}\otimes I)(|\Psi\rangle_{AB}\langle\Psi|)] \quad, \label{eq:coherence}
\end{equation}
where $\mathcal{E}$ represents the quantum operation of a noisy channel and
$\rho_{A}$ is the initial input state, $|\Psi\rangle_{AB}$ is the purified state of $\rho_{A}$, i.e., $\rho_{A}=\text{Tr}_{B}(|\Psi\rangle_{AB}\langle\Psi|)$, and $I$ is the identity operator. $S\left[ \rho  \right] =  - {\rm Tr}\left( {\rho \log \rho } \right)$ is the von Neumann entropy of a density matrix $\rho$.
Here the term channel is equally applicable to either a single fiber, or the combined fibers in the experimental setup. The quantum capacity
\begin{equation}\label{Q_1}
\mathcal{Q}_{1} \equiv \mathop {\max }\limits_\rho  \, {{\cal I}_c}
\end{equation}
of a single-use quantum channel is defined by the maximum coherent information that can be achieved over all possible input states $\rho$. More generally, quantum capacity is defined through the average coherent information transmitted through multiple uses~\cite{Lloyd1997,{Shor2002},Devetak2005}; but for dephasing channels~\cite{Devetak2005_2}, both definitions are equivalent.

For any state $\rho$ in the two-dimension space, it can always be expressed in some orthogonal basis $\{|\psi\rangle,|\psi_{\perp}\}$ as
\begin{equation}\label{rho_3p}
\rho=\lambda_{0}|\psi\rangle\langle\psi|+\lambda_{1}|\psi_{\perp}\rangle\langle\psi_{\perp}| \quad,
\end{equation}
where $|\psi\rangle=\cos\theta|0\rangle+\sin\theta e^{i\phi}|1\rangle$ and $|\psi_{\perp}\rangle=\sin\theta|0\rangle-\cos\theta e^{i\phi}|1\rangle$, and $\lambda_{0}+\lambda_{1}=1$.
The corresponding purified state can be written as $|\Psi\rangle=\sqrt{\lambda_{0}}|\psi\rangle|0\rangle+\sqrt{\lambda_{1}}|\psi_{\perp}\rangle|1\rangle$. Therefore, the coherent information $\mathcal{I}_c$ for any quantum channel transmitting qubits depends only on three independent parameters, namely $\lambda_{0}$, $\theta$ and $\phi$, which are employed for describing the experimental data.

\begin{figure*}[t]
\begin{center}
\includegraphics [width= 5.7in]{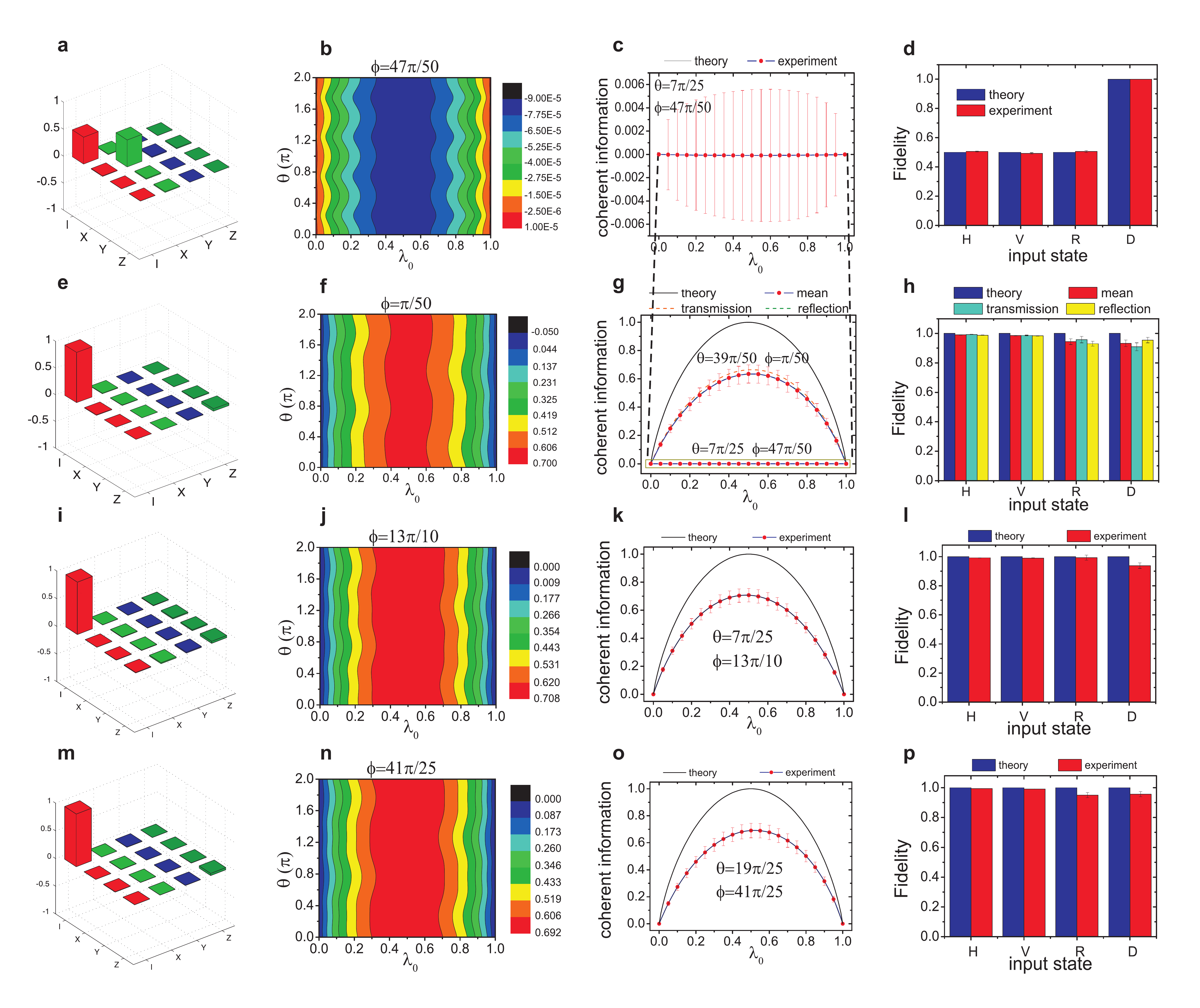}
\end{center}
\caption{(Color online). Experimental Results. {\bf a}, The real part of the experimental $\chi^{(s)}$-matrix for a single fiber. {\bf b}, The coherent information $\mathcal{I}_c$ calculated from $\chi^{(s)}$ by scanning $\theta$ and $\lambda_{0}$, with $\phi =47\pi/50$. {\bf c}, Coherent information of the single fiber as a function of $\lambda_{0}$ with $\theta=7\pi/25$ and $\phi=47\pi/50$.
{\bf d}, The fidelities of different states passing through the single fiber.
{\bf e-h}, The corresponding experimental results for a pair PM fibers with the HWPs in the interferometer setting to be $22.5^{\circ}$. {\bf i-l} and {\bf m-p}, The corresponding experimental results for the bi-directional use from left to right and right to left, respectively.
Error bars are estimated from standard deviation.} \label{fig:combine}
\end{figure*}

In order to fully characterize the action of a channel on the input signals $\rho$, we experimentally performed a full quantum process tomography~\cite{Chuang1997,Poyatos1997} on the quantum channels. By expanding the output state $\mathcal{E} (\rho)$ with a complete set of basis $\hat{E}_{m}$ of the Pauli operators $\{I, X, Y, Z\}$, the operation of the quantum process can be expressed as
\begin{equation}\label{chi_matrix}
\mathcal{E} (\rho) =\sum_{mn}\chi_{mn}\hat{E}_{m}\rho{\hat{E}_{n}}^{\dag} \quad.
\end{equation}
In this representation, the 4-by-4 matrix $\chi$ completely and uniquely characterizes the physical process $\mathcal{E}$. The matrix elements $\chi_{mn}$'s can be constructed from experimental tomographic measurements~\cite{OBrien2004} of four input signals, namely $\{|H\rangle, |V\rangle,| {D} \rangle \equiv \frac{1}{\sqrt{2}}(|H\rangle{+}|V\rangle), |R\rangle\equiv \frac{1}{\sqrt{2}}(|H\rangle{-}i|V\rangle)\}$. The experimental results for characterizing a single fiber is shown in fig.~\ref{fig:combine}a-d. We find that, to a good approximation, the individually-applied fibers are zero-quantum-capacity channels, i.e. does not allow quantum information transfer~\cite{SM}.

{\em Main experimental results --- } Denote the $\chi$-matrix for the unidirectional mode (fig.~\ref{fig:setup}b) as $\chi^{(m)}$. The output photon emerges in one of the exits, namely transmission and reflection port. Due to linearity, $\chi^{(m)}=p_T \chi^{(T)}+p_R \chi^{(R)}$ can be determined by the contributions, $\chi^{(T)}$ and $\chi^{(R)}$, from the transmission and reflection ports respectively. Here $p_{T}=0.498\pm0.004$ ($p_{R}=0.502\pm0.004$) is the experimentally-determined probability of detecting the signal photon at the transmission (reflection) port. Fig.~\ref{fig:combine}e shows the real part of the experimentally determined matrix elements of $\chi^{(m)}$ (the imaginary part is negligible and is shown in~\cite{SM}). It closely agrees with the theoretical prediction as an identity operation (i.e., perfect state transfer ${\cal E}\left( \rho  \right) = \rho$ for all $\rho$'s); the fidelity is found to be about $94.04\pm0.02$\%.

The quantum capacity $\mathcal{Q}_{1}$, determined by the maximal coherent information $\mathcal{I}_{c}$ calculated from $\chi^{(m)}$, is found to be $0.636\pm0.005$, which is obtained by setting $\lambda_{0}=0.52$, $\theta=39\pi/50$ and $\phi=\pi/50$. Fig.~\ref{fig:combine}f shows the coherent information calculated from $\chi^{(m)}$ as a function of $\lambda_{0}$ and $\theta$, with $\phi=\pi/50$. We can see that the maximum values of coherent information is achieved at around $\lambda_{0}=0.5$. The dependence of the coherent information on $\lambda_{0}$ is shown in fig.~\ref{fig:combine}g with $\theta=39\pi/50$ and $\phi=\pi/50$. The black solid line represents the theoretical prediction (i.e., the case for perfectly correlated noises). Red dots and the blue line represent the experimental coherent information calculated from $\chi^{(m)}$. The dashed green line and orange line represent the experimental results calculated from $\chi^{(R)}$ and $\chi^{(T)}$, respectively (the dashed green line and the blue solid line nearly overlap and only the blue solid line can be seen). The case for the single fiber is further shown in the dark yellow pane for comparison and the magnified case is shown in fig.~\ref{fig:combine}c as denoted by the two dashed lines.

The fidelities of different states passing through the paired fibers are shown in fig.~\ref{fig:combine}h. The blue columns represent the theoretical predictions and the red, cyan and yellow columns represent the experimental results of the mean, transmission and reflection cases, respectively. Given the high value of the quantum capacity achieved (from $10^{-16}$ to $0.636$), we conclude that the protocol of the proposed method does allow optical fibers with zero-quantum-capacity to transmit quantum information.

The experimental data for the bidirectional modes is presented in a similar way in fig.~\ref{fig:combine}i-l for quantum information transmitting from left to right (A to B) in fig.~\ref{fig:setup}c and from right to left (B to A) in fig.~\ref{fig:combine}m-p. In particular, the fidelity of the $\chi$ matrix compare to the identity from A to B is obtained to be $94.68\pm0.01\%$ ($\chi^{(AB)}$) and the case from B to A is equal to $95.13\pm0.03\%$ ($\chi^{(BA)}$), which shows the ability of the setup for transmitting quantum information bidirectionally.

\begin{figure}[tbph]
\begin{center}
\includegraphics [width= 1 \columnwidth]{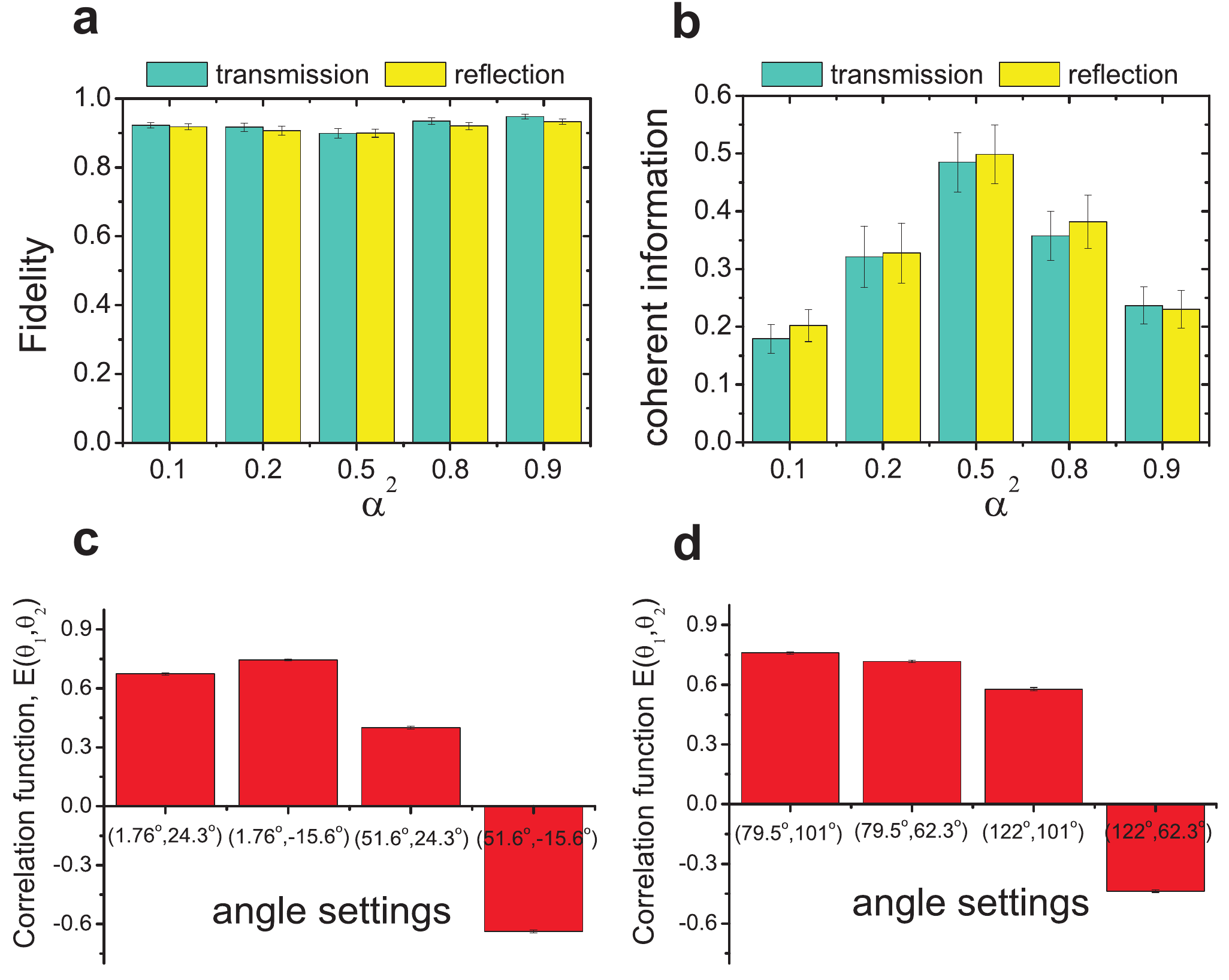}
\end{center}
\caption{(Color online). Experimental results for the entangled input states with one of the photons passing through the paired PM fibers. {\bf a}, The fidelities of the photon states with $\alpha^{2}=0.1, 0.2, 0.5, 0.8, 0.9$. {\bf b}, The coherent information with the corresponding input states. {\bf c}, and {\bf d}, represent the components for obtaining the CHSH values of the transmission ($S=2.457\pm0.025$) and reflection ($2.492\pm0.026$) ports, respectively. Error bars are estimated from standard deviation.} \label{fig:CHSH}
\end{figure}

{\em Application to testing quantum non-locailty--- }
Our setup can also be used to transmit nonlocal quantum information~\cite{Cavalcanti2011} or entangled photons~\cite{Huang2011}. We prepare different kinds of entangled states of the form $|\Psi\rangle=\alpha|HH\rangle+\beta|VV\rangle$ from two type-I beta-barium-borate (BBO) crystals \cite{Kwiat1999}, where $\alpha^{2}$ is set to be $\{0.1, 0.2, 0.5, 0.8, 0.9\}$. Our goal is to verify the nonlocality of the output state with maximal entanglement by testing the Clauser-Horne-Shimony-Holt (CHSH) inequality\cite{Clauser1969},
\begin{equation}
S=E(\theta_{1},\theta_{2})+E(\theta_{1},\theta_{2}')+E(\theta_{1}',\theta_{2})-E(\theta_{1}',\theta_{2}') \quad,
\end{equation}
where
\begin{equation}
E(\theta_{1},\theta_{2})=\frac{c(\theta_{1},\theta_{2})+c(\theta_{1}^{\perp},\theta_{2}^{\perp})
-c(\theta_{1},\theta_{2}^{\perp}-c(\theta_{1}^{\perp},\theta_{2}))}
{c(\theta_{1},\theta_{2})+c(\theta_{1}^{\perp},\theta_{2}^{\perp})
+c(\theta_{1},\theta_{2}^{\perp}+c(\theta_{1}^{\perp},\theta_{2}))},
\end{equation}
$\theta_{j}^{\perp}=\theta_{j}+90^{\circ}$, $j=1,2$ and
$c(\theta_{1},\theta_{2})$ is the coincidence counts with the
polarization angle settings $\theta_{1}$ in mode $A$ and
$\theta_{2}$ in mode $B$.

Fig.~ \ref{fig:CHSH}a and b characterize the properties of the setup for transferring entangled states. Fig.~\ref{fig:CHSH}a shows the corresponding state fidelities at the transmission (cyan columns) and reflection (yellow columns) ports, respectively. The fidelities of all output states are larger than 90\%. The coherent information with the corresponding input states are further shown in the fig.~\ref{fig:CHSH}b. The output state $\rho_{A}'$ ($=\mathcal{E}(\rho_{A})$) is obtained by tracing the photon $B$ of the final two photon state $\rho_{AB}'$ ($=(\mathcal{E}\otimes I)(|\Psi\rangle_{AB}\langle\Psi|)$) in order to calculate the coherent information. Cyan columns and yellow columns represent the experimental results at the transmission and reflection ports, respectively. We can see that the maximal coherent information is achieved with the setting of $\alpha^{2}=0.5$, which agree with the previous results.

Fig. \ref{fig:CHSH}c and d shows the experimental results of the correlations of the transmitted and the reflected states, respectively. The angles settings are calculated from the final density matrixes to maximize the values of $S$ ($S\leq2$ for any local realistic theory). In our experiment, we obtain the $S=2.457\pm0.025$ for fig. \ref{fig:CHSH}c and $S=2.492\pm0.026$ for fig. \ref{fig:CHSH}d. Both results violate the classical limit well above experimental errors (over about $18$ standard deviations), indicating the incompatibility of local realistic theories.

In summary, this new method allows us to use a pair of PM fibers, each of which has almost zero quantum capacity, for transferring quantum information. The setup can be used to transmit quantum information bidirectionally. When applied to entanglement distribution, the ability to preserve nonlocality promises practical applications for securing long-distance quantum communications.  Furthermore, our experimental setting provides a new physical platform for enriching the structure of the quantum channel theory when applied to correlated channels and the setup has potential application for protecting quantum information in large-scale photonics implementation of quantum algorithms.

We thank Y. Ouyang, J. Fitzsimons and K. Li for valuable comments and suggestions. This work was supported by the National Basic Research Program of China (Grants No. 2011CB9212000), National Natural Science Foundation of China (Grant Nos. 11274297, 11004185, 60921091, 11274289, 10874170), the Fundamental Research Funds for the Central Universities (Grant No. WK 2030020019), Program for New Century Excellent Talents in University, Science foundation for the excellent PHD thesis (Grant Nos. 201218) and the CAS.

%
%
%
%
%
%
%

\clearpage

\onecolumngrid


\pagestyle{empty}

\section{Supplemental Material}

\tableofcontents

\newpage

\section{Theoretical details}

\subsection{Dephasing property of a single PM Fiber}
We first consider a single PM fiber with the bases setting at the horizontal ($|H\rangle$) and vertical ($|V\rangle$) directions, which means that for the fiber length involved in the experiment, the horizontal and vertical polarization does not mix, i.e.,
\begin{equation}
{\cal E}\left( {\left|  H  \right\rangle \left\langle  H  \right|} \right) = \left|  H  \right\rangle \left\langle  H  \right| \, {, }  \,
{\cal E}\left( {\left|  V  \right\rangle \left\langle  V  \right|} \right) = \left|  V  \right\rangle \left\langle  V  \right| \quad.
\end{equation}
where ${\cal E}$ represents the operation of the PM fiber. Now, consider a general single photon state initialized as
\begin{equation}
 \int {d\omega } g\left( \omega  \right)\left( {\alpha \left| H \right\rangle  + \beta \left| V \right\rangle } \right)\left| \omega  \right\rangle \quad,
\end{equation}
where a qubit of information is encoded into the polarization degrees of freedom (DOF) $\{|H\rangle,|V\rangle\}$. Here we assume that the initial state is decoupled from the frequency DOF $\left| \omega  \right\rangle $ with the complex amplitude $g(\omega)$, which from the point of view of quantum channel theory, can be considered as part of the environment.

We first consider the component of a photon state with frequency $\omega$ passing through a single PM fiber. The dephasing mechanism induce a frequency-dependent phase between the two components of the photon (a global phase is omitted) \cite{Kumar2011},
\begin{equation}\label{app:phased_state}
\alpha \left| H \right\rangle  + \beta \left| V \right\rangle  \to \alpha \left| H \right\rangle  + \beta {e^{i\tau \left( \omega  \right)}}\left| V \right\rangle  \quad,
\end{equation}
where
\begin{equation}
\tau (\omega)=\kappa\omega
\end{equation}
and
\begin{equation}\label{app:kappa}
\kappa=L\Delta n/c \quad.
\end{equation}
Here $\Delta n$ is the difference between the refractive indexes of two polarizations along the basis of the PM fiber. $L$ is the length of the PM fiber and $c$ is the speed of the light in vacuum. The corresponding reduced density matrix is
\begin{equation}
\left( {\begin{array}{*{20}{c}}
{{{\left| \alpha  \right|}^2}}&{\alpha {\beta ^*}{e^{ - i\tau \left( \omega  \right)}}}\\
{{\alpha ^*}\beta {e^{i\tau \left( \omega  \right)}}}&{{{\left| \beta  \right|}^2}}
\end{array}} \right) \quad.
\end{equation}

For a Gaussian function like frequency distribution
\begin{equation}
f\left( \omega  \right) = {\left| {g\left( \omega  \right)} \right|^2} = \frac{2}{\sqrt{\pi}\sigma}\exp(-\frac{4(\omega-\omega_{0})^{2}}{\sigma^{2}})
\end{equation}
of the photon with the center frequency $\omega_{0}$ and a width $\sigma$, the value of the off-diagonal element
\begin{equation}
\exp(-\frac{1}{16}\kappa^{2}\sigma^{2}+i\kappa\omega_{0}) \quad,
\end{equation}
of the final density matrix decays rapidly with $L$. When $L$ is sufficiently long, the off-diagonal elements vanish and the coherence in the basis of the fiber is completely destroyed. In our case, the coherence length of the photon calculated by
\begin{equation}
\Delta L=\lambda^{2}/\Delta \lambda
\end{equation}
is about 213 $\mu$m, where $\lambda=800$ nm is the center wavelength of the photon and the width $\Delta \lambda=3$ nm is defined by the interference filter. As a result, when the length of the PM fiber reaches about $L=0.61$ m, the coherence of the photon is completely destroyed, where
\begin{equation}
L=\Delta L/\Delta n
\end{equation}
and $\Delta n=3.5\times10^{-4}$ (Nufern PM780-HP).

\subsection{The underlying design principle of the approach}
In our experiment, we use the polarization beam splitters (PBS) to encode and decode two PM fibers with the basis setting at $\{|H\rangle,|V\rangle\}$ (see fig. 1 in the main text). The PBS behaves like a CNOT gate, with the path degrees of freedom $\{|0\rangle,|1\rangle\}$ controlled by the polarization, where $|0\rangle$ and $|1\rangle$ represent the transmission and reflection cases. More precisely,
\begin{eqnarray}
\left| H \right\rangle \left| 0 \right\rangle  &\to& \left| H \right\rangle \left| 0 \right\rangle \quad ,\\
\left| H \right\rangle \left| 1 \right\rangle  &\to& \left| H \right\rangle \left| 1 \right\rangle \quad, \\
\left| V \right\rangle \left| 0 \right\rangle  &\to& \left| V \right\rangle \left| 1 \right\rangle \quad ,\\
\left| V \right\rangle \left| 1 \right\rangle  &\to& \left| V \right\rangle \left| 0 \right\rangle \quad .
\end{eqnarray}
Different types of setting of the half-wave plates (HWP) in the interferometer allow our setup to be operated in the unidirectional (fig. 1b in the main text) or bidirectional (fig. 1c in the main text). The efficiencies of both modes are analyzed below.

{\em Unidirectional mode---} Consider a general initial photon state,
\begin{equation}
|\phi_{in} \rangle=\alpha|H\rangle+\beta|V\rangle
\end{equation}
with a frequency $\omega$. $\alpha$ and $\beta$ are some unknown complex numbers and $|\alpha|^2+|\beta|^2=1$. After passing through the encoded PBS, the state becomes entangled with the path DOF,
\begin{equation}
|\phi_{ph}\rangle=\alpha|H\rangle |0\rangle+\beta|V\rangle |1\rangle \quad,
\end{equation}
where the horizontal component is transmitted ($|0\rangle$) and the vertical component is reflected ($|1\rangle$). Next, HWPs are applied to each of the path (HWP1 and HWP3), which rotate the horizontal state $\left| H \right\rangle$ to the diagonal state $\left|  D  \right\rangle  = \left( {\left| H \right\rangle  + \left| V \right\rangle } \right)/\sqrt 2$, and $\left| V \right\rangle$ to the anti-diagonal state $\left|  J  \right\rangle  = \left( {\left| H \right\rangle  - \left| V \right\rangle } \right)/\sqrt 2$. The state becomes
\begin{equation}
\left| {{\phi _{ph}}} \right\rangle  = \alpha \left|  D  \right\rangle \left| 0 \right\rangle  + \beta \left|  J  \right\rangle \left| 1 \right\rangle \quad.
\end{equation}

After passing through the corresponding PM fibers, the photon state becomes
\begin{equation}
\frac{\alpha}{\sqrt{2}} ( {\left| H \right\rangle  + {e^{i{\tau _1}\left( \omega  \right)}}\left| V \right\rangle } ) \left| 0 \right\rangle  + \frac{\beta}{\sqrt{2}}( {\left| H \right\rangle  - {e^{i{\tau _2}\left( \omega  \right)}}\left| V \right\rangle } )\left| 1 \right\rangle \, .
\end{equation}
Then, we apply two other HWPs (HWP2 and HWP4) with the same angles set as in the encoding stage, and get

\begin{eqnarray}
|\phi_{ph}\rangle & = & \frac{\alpha}{{2}} [(1+e^{i\tau_1(\omega)}) |H \rangle |0\rangle+ (1-e^{i\tau_1(\omega)}) |V \rangle |0\rangle] \nonumber \\
                  & +& \frac{\beta}{{2}} [(1-e^{i\tau_2 (\omega)}) |H \rangle |1\rangle + (1+e^{i\tau_2(\omega)}) |V \rangle |1\rangle].
\end{eqnarray}

By implementing the decoding PBS which transmits horizontal component and reflects the vertical component, the state becomes
\begin{eqnarray}
|\phi_{ph}\rangle & = & \frac{\alpha}{{2}}[(1+e^{i\tau_1 (\omega)}) |H \rangle |0\rangle+(1-e^{i\tau_1(\omega)}) |V\rangle |1\rangle] \nonumber \\
& + & \frac{\beta}{{2}}[(1-e^{i\tau_2(\omega)}) |H \rangle |1\rangle +(1+e^{i\tau_2(\omega)}) |V \rangle |0\rangle] \, .
\end{eqnarray}

Now, we apply a HWP with the angle setting at $45^{\circ}$ at reflection port in figure 1, which flips $|H\rangle$ and $|V\rangle$ ($|H\rangle\Leftrightarrow|V\rangle$). Consequently, the final state becomes,
\begin{eqnarray}\label{app:output_asymmetry}
\left| {{\phi _{ph}}} \right\rangle  = \frac{1}{2}( {\alpha ( {1 + {e^{i{\tau _1}( \omega  )}}} )\left| H \right\rangle  + \beta ( {1 + {e^{i{\tau _2}( \omega  )}}} )| V \rangle })| 0 \rangle \nonumber  \\
 + \frac{1}{2}( {\alpha ( {1 - {e^{i{\tau _1}( \omega  )}}} )| H \rangle  + \beta ( {1 - {e^{i{\tau _2}( \omega  )}}} )| V \rangle } )| 1 \rangle.
\end{eqnarray}

The corresponding reduced density matrix of the system containing the polarization DOF has the following form:
\begin{equation}
\left( {\begin{array}{*{20}{c}}
{{{\left| \alpha  \right|}^2}} & {\alpha {\beta ^*}  \eta \left( {\Delta \tau } \right)}\\
{{\alpha ^*}\beta \eta^* \left( {\Delta \tau } \right) }&{{{\left| \beta  \right|}^2}}
\end{array}} \right) \quad,
\end{equation}
where
\begin{equation}
\eta \left( {\Delta \tau } \right) \equiv {e^{i\Delta \tau \left( \omega  \right)/2}}\cos \left( {\Delta \tau \left( \omega  \right)/2} \right) \quad,
\end{equation}
and
\begin{equation}
\Delta \tau  \equiv {\tau _1} - {\tau _2} = \left( {{\kappa _1} - {\kappa _2}} \right)\omega \quad,
\end{equation}
where $\kappa_1$ and  $\kappa_2$ (cf. Eq.~(\ref{app:kappa})) contains the material properties of the PM fibers. When we sum over all frequencies in a boardband distribution $f(\omega)$, the coherence factor decays as
\begin{equation}
\Gamma  \equiv \int {} f\left( \omega  \right)\eta \left( {\Delta \tau } \right)d\omega \quad.
\end{equation}

We therefore see that when
\begin{equation}
{\tau _1} \approx {\tau _2} \equiv \tau \quad,
\end{equation}
the output state is a pure state for any frequency, and is identical with the input state, i.e., perfect state transfer.

In the idealized case where the two channels are perfectly correlated, we can write Eq.~(\ref{app:output_asymmetry}) as
\begin{equation}
\left| {{\phi _{ph}}} \right\rangle  = \frac{1}{2}( {1 + {e^{i\tau \left( \omega  \right)}}} )\left| {{\phi _{in}}} \right\rangle \left| 0 \right\rangle  + \frac{1}{2}( {1 - {e^{i\tau \left( \omega  \right)}}} )\left| {{\phi _{in}}} \right\rangle \left| 1 \right\rangle \, ,
\end{equation}
which shows that the initial state reliably transmitted in either of the output.

{\em Bidirectional mode---} The physical setting for the bidirectional mode is shown in fig. 1c in the main text. In such case, the angles of HWP1 and HWP2 are set to be $0^{\circ}$, and HWP3 and HWP4 are set to be $45^{\circ}$.

When the quantum state, $|\phi_{in} \rangle$, is sent from the A side to B side (from left to right), the PBS entangles the path DOF, and the state becomes
\begin{equation}
|\phi_{ph}\rangle=\alpha|H\rangle |0\rangle+\beta|V\rangle |1\rangle \quad,
\end{equation}
The horizontal polarization along the transmitted path $(\left| 0 \right\rangle )$ is unchanged, but the vertical polarization at the reflected path $(\left| 1 \right\rangle )$  flips to the horizontal state. The information of the input quantum state is then transferred to the path DOF,
\begin{equation}
|\phi_{ph}\rangle=|H\rangle \left( {\alpha \left| 0 \right\rangle  + \beta |\left| 1 \right\rangle } \right) \quad.
\end{equation}
After the photon passes through the pair of PM fibers, there are phase shifts ${{\tau _{H1,H2}}\left( \omega  \right)}$ induced from each path for each frequency component $\omega$,
\begin{equation}
|\phi_{ph}\rangle=\left| H \right\rangle \left( {\alpha {e^{i{\tau _{H1}}\left( \omega  \right)}}\left| 0 \right\rangle  + \beta {e^{i{\tau _{H2}}\left( \omega  \right)}}|\left| 1 \right\rangle } \right) \quad,
\end{equation}
Then, HWP4 at the reflected path flips the polarization back to $|V\rangle $, which gives the following state,
\begin{equation}
|\phi_{ph}\rangle=\alpha {e^{i{\tau _{H1}}\left( \omega  \right)}}\left| H \right\rangle \left| 0 \right\rangle  + \beta {e^{i{\tau _{H2}}\left( \omega  \right)}}\left| V \right\rangle |\left| 1 \right\rangle  \quad,
\end{equation}

Finally, the two paths cross at the decoding PBS, which separates the path DOF,
\begin{equation}
|\phi_{ph}\rangle=\left( {\alpha {e^{i{\tau _{H1}}\left( \omega  \right)}}\left| H \right\rangle  + \beta {e^{i{\tau _{H2}}\left( \omega  \right)}}\left| V \right\rangle } \right)\left| 0 \right\rangle  \quad,
\end{equation}

The corresponding reduced density matrix of the polarization state has the following form:
\begin{equation}
\left( {\begin{array}{*{20}{c}}
{{{\left| \alpha  \right|}^2}} & {\alpha {\beta ^*}  e^{i {\Delta \tau_{H} }}}\\
{{\alpha ^*}\beta e^{-i {\Delta \tau_{H} }} }&{{{\left| \beta  \right|}^2}}
\end{array}} \right) \quad,
\end{equation}
with $\Delta \tau_{H}=\tau_{H1}-\tau_{H2}$. Whenever the errors are correlated, i.e., ${\tau _{H1}} \approx {\tau _{H2}} \equiv \tau_{H}$, the signal photon with the state identical to the initial one can be detected along the path $\left| 0 \right\rangle$. The phase difference $\Delta \tau_{H}$ causes dephasing to the output state as expected.

In a similar way, quantum information can be sent from the B side to A side (from right to left) as indicated in fig. 1c. The decoded (encoded) PBS for the case of A to B is now treated as the encoded (decoded) PBS. Consider another general initial photon state $|\psi_{in}\rangle=\gamma|H\rangle+\delta|V\rangle$ ($\gamma$ and $\delta$ are some unknown complex numbers and $|\gamma|^2+|\delta|^2=1$). After passing the encoded PBS, the state becomes
\begin{equation}
|\psi_{ph}\rangle=\gamma|H\rangle |0\rangle+\delta|V\rangle |1\rangle \quad.
\end{equation}
HWP4 located at the transmitted path $(\left| 0 \right\rangle )$ change $|H\rangle$ to $-|V\rangle$, and HWP2 at the reflected path $(\left| 1 \right\rangle )$ changes $|V\rangle$ to  $-|V\rangle$. The information of the input quantum state is transferred to the path DOF,
\begin{equation}
|\psi_{ph}\rangle=-|V\rangle \left( {\gamma \left| 0 \right\rangle  + \delta |\left| 1 \right\rangle } \right) \quad,
\end{equation}
After the photon passes through the pair of PM fibers, there are phase shifts ${{\tau _{V1,V2}}\left( \omega  \right)}$ induced from each path for each frequency component $\omega$,
\begin{equation}
|\psi_{ph}\rangle=-\left| V \right\rangle \left( {\gamma {e^{i{\tau _{V1}}\left( \omega  \right)}}\left| 0 \right\rangle  + \delta {e^{i{\tau _{V2}}\left( \omega  \right)}}|\left| 1 \right\rangle } \right) \quad,
\end{equation}
Then, HWP3 at the transmitted path changes $-|V\rangle$ to $|H\rangle $ and HWP1 at the reflected path change $-|V\rangle$ to $|V\rangle$, which gives the following state,
\begin{equation}
|\psi_{ph}\rangle=\gamma {e^{i{\tau _{V1}}\left( \omega  \right)}}\left| H \right\rangle \left| 0 \right\rangle  + \delta {e^{i{\tau _{V2}}\left( \omega  \right)}}\left| V \right\rangle |\left| 1 \right\rangle  \quad,
\end{equation}

Finally, by implementing the decoding PBS which transmits horizontal component and reflects the vertical component, the state becomes,
\begin{equation}
|\psi_{ph}\rangle=\left( {\gamma {e^{i{\tau _{V1}}\left( \omega  \right)}}\left| H \right\rangle  + \delta {e^{i{\tau _{V2}}\left( \omega  \right)}}\left| V \right\rangle } \right)\left| 0 \right\rangle  \quad,
\end{equation}

The corresponding reduced density matrix of the system has the following form:
\begin{equation}
\left( {\begin{array}{*{20}{c}}
{{{\left| \gamma  \right|}^2}} & {\gamma {\delta ^*}  e^{i {\Delta \tau_{V} }}}\\
{{\gamma ^*}\delta e^{-i {\Delta \tau_{V} }} }&{{{\left| \delta  \right|}^2}}
\end{array}} \right) \quad,
\end{equation}
with $\Delta \tau_{V} = \tau _{V1}- \tau _{V2}$. Whenever the errors are correlated, i.e., ${\tau _{V1}} \approx {\tau _{V2}} \equiv \tau$, the signal photon with the state identical to the initial one can be detected along the path $\left| 0 \right\rangle$. The phase difference $\Delta \tau_{V}$ causes dephasing to the output state as expected.

\subsection{Coherent Information}
A quantum channel can be defined by a completely-positive quantum operation $\mathcal{E}$, which acts on an input state $\rho^Q$ of a quantum system $Q$, such that the output state ${\rho ^{Q'}}$ of the quantum channel is given by

\begin{equation}
{\rho ^{Q'}} = {\cal E}\left( {{\rho ^Q}} \right) \quad.
\end{equation}

In principle the input state can be any state, mixed or pure state. In the case of mixed states, it can always get ``purified" and become a pure state $\left| {{\psi ^{QR_{f}}}} \right\rangle \left\langle {{\psi ^{QR_{f}}}} \right|$, where
\begin{equation}
\rho^{Q} = {\rm tr}_{R_{f}}\left( {\left| {{\psi ^{QR_{f}}}} \right\rangle \left\langle {{\psi ^{QR_{f}}}} \right|} \right) \quad,
\end{equation}
by introducing a reference (or imaginary) system $R_{f}$. The {\it entropy exchange} of~$\mathcal{E}$ with input state $\rho^Q$ is defined as~\cite{Nielsen2000}
\begin{equation}
S\left( {{\cal E},{\rho ^Q}} \right) \equiv S\left( {R_{f}',Q'} \right) =  - {\rm tr}( {{\rho ^{Q'R_{f}'}}\log ( {{\rho ^{Q'R_{f}'}}} )}) \quad.
\end{equation}
which can be considered as the entropy induced by the quantum operation $\mathcal{E}$.

Quantum {\it coherent information} is defined as~\cite{Schumacher1996}
\begin{equation}
\mathcal{I}_{c}(\mathcal{E},\rho^Q)=S[\mathcal{E}(\rho^Q)]-S[(\mathcal{E}\otimes I)(|\psi^{QR_{f}}\rangle\langle\psi^{QR_{f}}|)],
\end{equation}
where the second term is equal to the entropy exchange. Quantum coherent information can be regarded as the amount of quantum information that is
transmitted through the noisy channel.

One important property about coherent information is that it satisfies a quantum data processing inequality, which is an extension of the classical data processing inequality in the quantum domain. More precisely, consider two successive application of two same or different channels, $\mathcal{E}_1$ and $\mathcal{E}_2$, to some input state $\rho$ so that
\begin{equation}
\rho ' = \mathcal{E}_1\left( \rho  \right) \quad {\rm and} \quad \rho '' = \mathcal{E}_2\left( {\rho '} \right).
\end{equation}

The quantum data processing inequality states that~\cite{Schumacher1996}
\begin{equation}
S\left( \rho  \right) \ge \mathcal{I}_c\left( {{\mathcal{E}_1},\rho } \right) \ge \mathcal{I}_c\left( {{\mathcal{E}_2} \circ {\mathcal{E}_1},\rho } \right) \quad,
\end{equation}
which means that coherent information has the property that it cannot increase after each interaction with the environment. This result is similar to the role of mutual information in classical data processing inequality.

\subsection{Single-use Channel Capacity of the Completely Dephasing Channel}
Quantum capacity for a single-use channel is defined as the maximum coherent information the channel can result:
\begin{equation}
\mathcal{Q}_{1} \equiv \mathop {\max }\limits_{{\rho ^Q}}\mathcal{I}_{c}(\mathcal{E},\rho^Q).
\end{equation}
where recall that $\mathcal{I}_{c}(\mathcal{E},\rho^Q)=S[\mathcal{E}(\rho^Q)]-S[(\mathcal{E}\otimes I)(|\psi^{QR_{f}}\rangle\langle\psi^{QR_{f}}|)]$, $ | \psi^{QR} \rangle $ is the purified state of $\rho^Q$, and $I$ represents the identity operator. Here we consider the channel capacity of the completely depahsing channel, which is expected to be zero. We provide an explicit construction.

Consider a qubit state $\rho^Q$ that is purified with a reference system. We can always write the resulting state in terms of Schmidt decomposition:
\begin{equation}
|\Psi^Q\rangle=\sqrt{\lambda_{0}}|\psi\rangle|0\rangle_{r}+\sqrt{\lambda_{1}}|\psi_{\perp}\rangle|1\rangle_{r}.
\end{equation}
where
\begin{equation}
|\psi\rangle=\cos\theta|0\rangle+\sin\theta e^{i\phi}|1\rangle \quad,
\end{equation}
\begin{equation}
|\psi_{\perp}\rangle=\sin\theta|0\rangle-\cos\theta e^{i\phi}|1\rangle \quad,
\end{equation}
and $\lambda_{0}+\lambda_{1}=1$. $\{|0\rangle, |1\rangle\}$ and $\{|0\rangle_{r}, |1\rangle_{r}\}$ are the basis of the interested and reference systems (for the simplicity, we omitted the subscripts below). Note that there are three independent parameters, namely $\lambda_{0}$, $\theta$ and $\phi$, to characterize $\rho^{Q}$. Explicitly,
\begin{equation}
\rho^{Q}=\lambda_{0}|\psi\rangle\langle\psi|+\lambda_{1}|\psi_{\perp}\rangle\langle\psi_{\perp}|.
\end{equation}

First of all, by definition, the completely dephasing channel removes all off-diagonal elements in $\rho^Q$. Therefore, we have
\begin{equation}
{\cal E}\left( \rho  \right) = \left( {\begin{array}{*{20}{c}}
{{\lambda _0}{{\cos }^2}\theta  + {\lambda _1}{{\sin }^2}\theta }&0\\
0&{{\lambda _0}{{\sin }^2}\theta  + {\lambda _1}{{\cos }^2}\theta }
\end{array}} \right)
\end{equation}
Note that the entropy $S[\mathcal{E}(\rho)]$ is maximized when
\begin{equation}
\lambda_{0}=\lambda_{1}=1/2 \quad {\rm and} \quad \cos\theta=\sin\theta=1/\sqrt{2} \quad.
\end{equation}

Now let us consider another term. We first write explicitly
\begin{eqnarray*}
|\Psi^Q\rangle\langle\Psi^Q|&=&\lambda_{0}|\psi\rangle\langle\psi|\otimes|0\rangle\langle0|+\lambda_{1}|\psi_{\perp}\rangle\langle\psi_{\perp}|\otimes|1\rangle\langle1|\nonumber\\
                          &+&\sqrt{\lambda_{0}\lambda_{1}}(|\psi\rangle\langle\psi_{\perp}|\otimes|0\rangle\langle1|+|\psi_{\perp}\rangle\langle\psi|\otimes|1\rangle\langle0|).
\end{eqnarray*}
Applying the completely dephasing channel to the first qubit, we can see that
\begin{eqnarray*}
\mathcal{E}{\otimes} I |\Psi^Q\rangle\langle\Psi^Q| {=} \lambda_{0}({{\cos^{2}}{\theta}}|0\rangle\langle0| {+} {\sin^{2}}\theta|1\rangle\langle1|) {\otimes} {|0\rangle\langle0|} \nonumber \\
+ \lambda_{1}({\sin^{2}}\theta|0\rangle\langle0|{+}{\cos^{2}}\theta|1\rangle\langle1|) {\otimes} |1\rangle\langle1| \nonumber \\
+\sqrt{\lambda_{0}\lambda_{1}}\cos\theta\sin\theta(|0\rangle\langle0|-|1\rangle\langle1|){\otimes}(|1\rangle\langle0|+|0\rangle\langle1|).
\end{eqnarray*}

To calculate the entropy of the resulting state, we need to solve for the eigenvalues; we re-write into two block-diagonal subspaces
\begin{equation}
(\mathcal{E}\otimes I)(|\Psi\rangle\langle\Psi|) \equiv |0\rangle\langle0|\otimes M_{1}+|1\rangle\langle1|\otimes M_{2},
\end{equation}
where
$$
M_{1} \equiv \left( \begin{array}{cc}
\lambda_{0}\cos^{2}\theta & \sqrt{\lambda_{0}\lambda_{1}}\cos\theta\sin\theta \\
\sqrt{\lambda_{0}\lambda_{1}}\cos\theta\sin\theta & \lambda_{1}\sin^{2}\theta
\end{array} \right)
$$
and
$$
M_{2} \equiv \left( \begin{array}{cc}
\lambda_{0}\sin^{2}\theta & -\sqrt{\lambda_{0}\lambda_{1}}\cos\theta\sin\theta \\
-\sqrt{\lambda_{0}\lambda_{1}}\cos\theta\sin\theta & \lambda_{1}\cos^{2}\theta
\end{array} \right)
$$
Recall that the eigenvalues of real and symmetric $2\times2$ matrix
$\left( \begin{array}{cc}
a & c \\
c & b
\end{array} \right)$
are
\begin{equation}
\frac{1}{2}[(a+b)\pm\sqrt{(a-b)^{2}+4c^{2}}] \quad.
\end{equation}
We found that the eigenvalues of $M_{1}$ are
\begin{equation}
{{\lambda _0}{{\cos }^2}\theta  + {\lambda _1}{{\sin }^2}\theta } \quad {\rm and} \quad 0 \quad .
\end{equation}

Similarly, the eigenvalues of $M_{2}$ are found to be
\begin{equation}
\lambda_{0}\sin^{2}\theta+\lambda_{1}\cos^{2}\theta \quad {\rm and} \quad 0.
\end{equation}

So, there are only two non-zero eigenvalues, which are the same for ${\cal E}\left( \rho  \right)$ and $(\mathcal{E}\otimes I)(|\Psi^Q\rangle\langle\Psi^Q|)$. Hence, the coherent information is always equal to zero.

For larger entangled states, the quantum capacitiy is generalized to~\cite{Lloyd1997,Shor2002,Devetak2005}
\begin{equation}
\mathcal{Q} \equiv {\mathop {\lim} \limits_{n\rightarrow\infty}} \mathop{\max} \limits_{\rho} \, \mathcal{I}_{c}^{\otimes n} / n \quad.
\end{equation}
Since dephasing channel belongs to the class of degradable channels, which implies that $\mathcal{Q}_1=\mathcal{Q}$~\cite{Devetak2005_2}. Therefore as long as $\mathcal{Q}_1 = 0$, no quantum information can be transferred through the channels, no matter how many of them is used.

\section{Experimental details}

\subsection{Reconstruction of the density matrix $\chi$}

The standard quantum process tomography usually leads to an unphysical density matrix. We then follow the maximal-likelihood method~\cite{OBrien2004} to reconstructed a positive, Hermitian matrix $\chi$. The physical density $\chi$ is defined as
\begin{equation}
\chi \equiv \hat{T}T/\text{Tr}(\hat{T}T) \quad ,
\end{equation}
where $T$ is related to 16 parameters~\cite{James2001}
\begin{equation}
T=\left( {\begin{array}{{cccc}}
{t_{1}}&{0}&{0}&{0}\\
{t_{5}+it_{6}}&{t_{2}}&{0}&{0}\\
{t_{11}+it_{12}}&{t_{7}+it_{8}}&{t_{3}}&{0}\\
{t_{15}+it_{16}}&{t_{13}+it_{14}}&{t_{9}+it_{10}}&{t_{4}}
\end{array}} \right) \quad.
\end{equation}

These parameters are obtained by minimizing the function
\begin{equation}
f( {\vec{t}} )=\sum_{i,j=1}^{4}\frac{1}{C}[c_{ij}-C\sum_{m,n=1}^{4}\langle \varphi_{j} | \hat{E_{m}} |\varphi_{i}\rangle
\langle\varphi_{j}|\hat{E_{n}}|\varphi_{i}\rangle\chi_{mn}(\vec{t})]^{2},
\end{equation}
where $C$ is the total number of counting photons, $c_{ij}$ is the corresponding counting with the input state $\varphi_{i}$ and the measurement state $\varphi_{j}$, and
\begin{equation}
\varphi_{i}\in\{H (\varphi_{1}), V (\varphi_{2}), D=\frac{1}{\sqrt{2}}(H+V) (\varphi_{3}), R=\frac{1}{\sqrt{2}}(H-iV) (\varphi_{4})\} \quad.
\end{equation}
The density matrix $\chi$ is reconstructed with the obtained 16 parameters $\{t_{1}, ......,t_{16}\}$. In our experiment, the initial density matrix is set to be normalized for the linear mapping of a quantum channel corresponds to a quantum state of a larger dimension~\cite{Jamiolkowski1972}. Compared with the minimization function in Ref.~\cite{OBrien2004}
, the parameter $\lambda$ is set to be zero in our case. The fidelity between the experimental result ($\chi^{(e)}$) and theoretical prediction ($\chi^{(t)}$) is calculated as $(\text{Tr}\sqrt{\sqrt{\chi^{(e)}}\chi^{(t)}\sqrt{\chi^{(e)}}})^{2}$.

With the experimentally determined $\chi$-matrix, we systematically searched for the maximum value for the coherent information $\mathcal{I}_c$; the smallest steps during the calculation are 0.01 for $\lambda_{0}$ and $\pi/50$ for $\theta$ and $\phi$, respectively.

\subsection{Features of the experimental results on a single-fiber characterization}

Fig.~2a-d in the main text shows the experimental results for a single-fiber characterization. Fig.~2a shows the experimental real part of the $\chi$-matrix, denoted as $\chi^{(s)}$, of the single PM fiber (the corresponding imaginary part is small , see the next section). Note that the two nearly-equal distributions of $I$ and $X$ indicate the strong dephasing effect of the fiber on the basis of $|D\rangle$ and $|J\rangle$
($\equiv \frac{1}{\sqrt{2}}(|H\rangle{-}|V\rangle)$), i.e., ${\cal E}\left( \rho  \right) \approx \left( {I\rho I + X\rho X} \right)/2 \equiv {\chi ^{(i)}}\left( \rho  \right)$. The fidelity of the experimental result $\chi^{(s)}$ is about~$99.86\pm0.01 \%$, which is calculated from the fidelity $(\text{Tr} \sqrt{\sqrt{\chi^{(s)}}\chi^{(i)}\sqrt{\chi^{(s)}}})^{2}$~\cite{OBrien2004}.
With the experimentally determined $\chi$-matrix, we systematically searched for the maximum value for the coherent information $\mathcal{I}_c$. We found that the maximal value of coherence information is about $8.55\times10^{-16} \pm 3.50\times10^{-16}$ with $\lambda_{0}=0$, $\theta=7\pi/25$ and $\phi=47\pi/50$.
Fig.~2b shows the coherence information calculated from $\chi^{(s)}$ by scanning $\theta$ and $\lambda_{0}$ with $\phi$ setting to be $47\pi/50$. We can see that the maximum value of coherence information is achieved with $\lambda_{0}$ equals to 0. The coherence information of the single fiber is further shown as a function of $\lambda_{0}$ with $\theta=7\pi/25$ and $\phi=47\pi/50$ (fig.~2c). The theoretical prediction (black line) agrees with the experimental result (blue line and red dots) which equals nearly to zero (the black line and the blue line nearly overlap and only the blue line can be seen). Error bars are estimated from standard deviation (see the section of error estimation) and the maximum deviation is about 0.006. The theoretical (blue columns) and experimental (red columns) fidelities of the four states $\{|H\rangle, |V\rangle, |D\rangle, |R\rangle\}$) are further compared in the fig.~2d. We find that they are in good agreement. Therefore, we conclude that the quantum capacity of the PM fiber is a good approximation to a zero-capacity quantum channel.

\subsection{Imaginary part of the density matrix $\chi$}

\begin{figure}[tbph]
\begin{center}
\includegraphics [width= 0.6 \columnwidth]{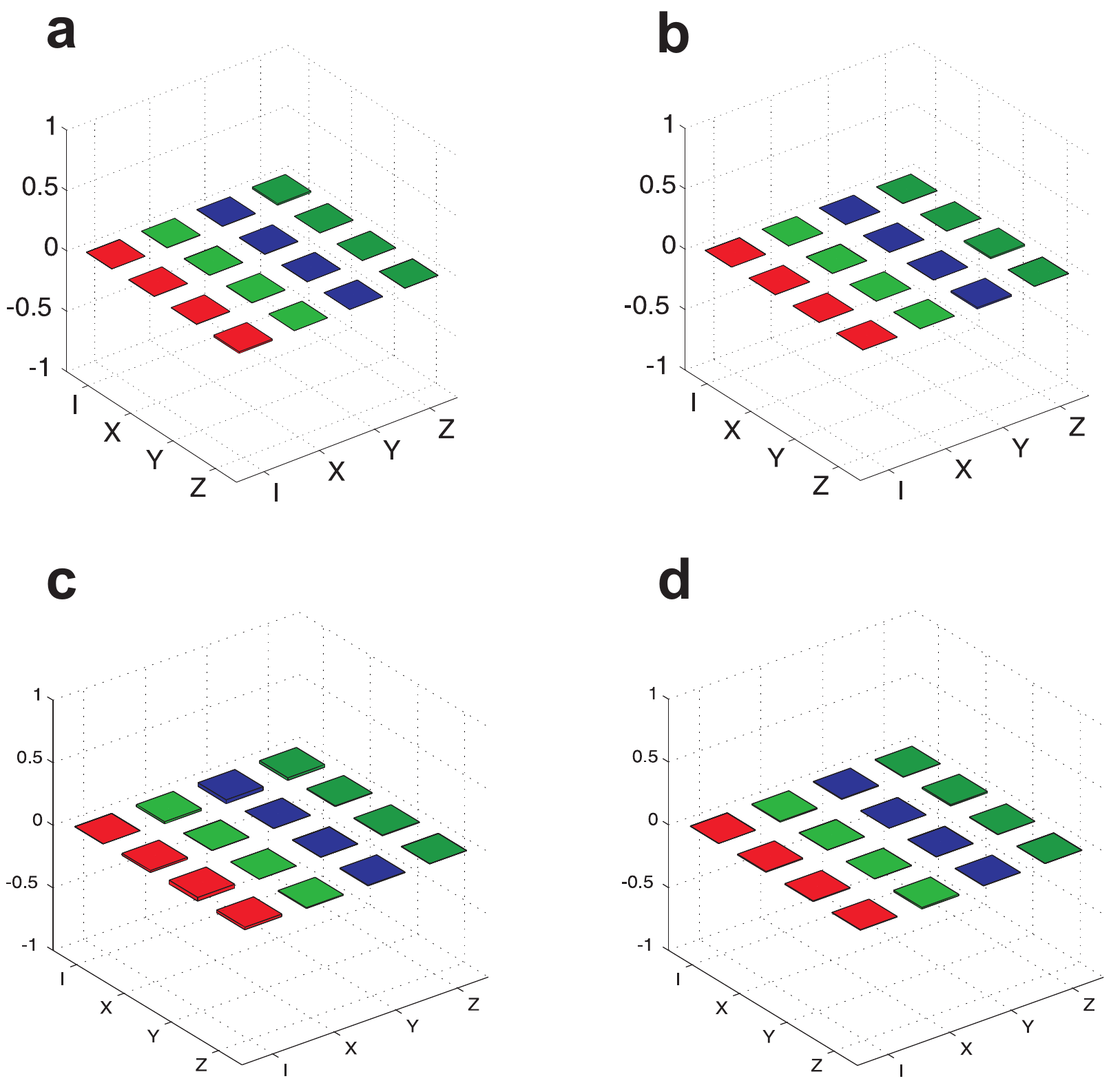}
\end{center}
\caption{(Color online). Experimental result for the imaginary parts of the density matrixes $\chi^{(s)}$ ({\bf a}), $\chi^{(m)}$ ({\bf b}), $\chi^{(AB)}$ (bidirectional mode from A to B, {\bf c}) and $\chi^{(BA)}$ (bidirectional mode from B to A, {\bf d}).} \label{fig:imaginary}
\end{figure}

Fig.~\ref{fig:imaginary} shows the experimental result for the imaginary parts of the density matrixes $\chi^{(s)}$ ({\bf a}), $\chi^{(m)}$ ({\bf b}), $\chi^{(AB)}$ (bidirectional mode from A to B, {\bf c}) and $\chi^{(BA)}$ (bidirectional mode from B to A, {\bf d}). We can see that all the elements are small and nearly equal to zero, which agree well with the theoretical predictions.

\subsection{Calibration and measurement}

Our experimental setup is shown in Fig.1 (main text). Ultraviolet (UV) pluses are used to pump two type-I beta-barium-borate (BBO) crystals to produce polarization-entangled photon pairs~\cite{Kwiat1999}. The ultraviolet pulses are frequency doubled from a mode-locked Ti:sapphire laser centered at 800 nm with
130 fs pulse width and 76 MHz repetition rate.

After compensating the birefringence effect between $H$ and $V$ in BBO crystals with quartz plates (CP), maximally entangled photon pairs emit into pathes $A$ and $B$. The photon in path $A$ passes through the quantum channel (either a single PM fiber or the encoded paired PM fibers) and sends to Bob. The photon in path $B$ is triggered into $| H \rangle$ and the photon in path $A$ prepared to the corresponding states ($\{|H\rangle, |V\rangle, |D\rangle, |R\rangle\}$) is sent to the interferometer (from left to right) to implement quantum process tomography.
For verifying the bidirectional use, the photon in path $A$ is then trigged into $| H \rangle$ and the photon in path $B$ prepared to the corresponding states ($\{|H\rangle, |V\rangle, |D\rangle, |R\rangle\}$) is sent to the interferometer (from right to left) to implement quantum process tomography. A feedback control system is used to lock the relative phase between the two arms of the interferometer (not shown in fig.1).

The polarization of the final state is then analyzed by a quarter-wave plate (QWP), a half-wave plate (HWP) and a polarization beam splitter (PBS) in each arms. The photons are detected by single photon avalanche detectors (SPADs) with 3 nm interference filters (IFs) in front of them. \\

\subsection{Error estimation}
In our experiment, the counting of each measurement is assumed to follow the Poisson distribution (the used subprogram is PoissonDistribution in Wolfram Mathematica 7.0). We randomly regroup 50 counting sets for each measurement quantity from the distribution countings. The values of the quantity can be calculated from the corresponding counting sets and 50 values are obtained. The error of the quantity is then estimated by the square root of the variance of the 50 values.

%
%
%
%
%
%
%
%
%
%
%
%
%
%


\begin{thebibliography}{99}
\bibitem{comment}
Here ``unidirectional" means one input from one side and one output at the other side of the communication channel. ``Bidirectional" means one input and one output are both located at each side. (Compare Fig.~\ref{fig:setup}b \& c)
\end{thebibliography}
\end{document}